\documentclass[english,PRE,twocolumn]{revtex4-1}
\usepackage[T1]{fontenc}
\usepackage[latin9]{inputenc}
\usepackage{amstext}
\usepackage{amssymb}
\usepackage{graphicx}
\usepackage{babel}
\begin{document}

\title{Neutral Aggregation in Finite Length Genotype space.}

\author{Bahram Houchmandzadeh}

\affiliation{CNRS, LIPHY, F-38000 Grenoble, France~\\
Univ. Grenoble Alpes, LIPHY, F-38000 Grenoble, France}
\begin{abstract}
The advent of modern genome sequencing techniques allows for a more
stringent test of the neutrality hypothesis of Darwinian evolution,
where all individuals have the same fitness. Using the individual
based model of Wright and Fisher, we compute the amplitude of \emph{neutral}
aggregation in the genome space, \emph{i.e., }the probability of finding
two individuals at genetic (hamming) distance $k$ as a function of
genome size $L$, population size $N$ and mutation probability per
base $\nu$. In well mixed populations, we show that for $N\nu<1/L$,
neutral aggregation is the dominant force and most individuals are
found at short genetic distances from each other. For $N\nu>1$ on
the contrary, individuals are randomly dispersed in genome space.
The results are extended to geographically dispersed population, where
the controlling parameter is shown to be a combination of mutation
and migration probability. The theory we develop can be used to test
the neutrality hypothesis in various ecological and evolutionary systems.
\end{abstract}
\maketitle

\section{Introduction.}

Aggregation of individuals is a common observation in evolutionary
and ecological systems. By aggregation, we mean the observation that
some areas of \emph{space} contain large numbers of individuals while
other parts contain relatively few. To be more precise, the variance
of population distribution is much larger than its mean. Consider
for example ecological communities where observations are made in
the \emph{real} space. Since the seminal work of Taylor et al.,\cite{Taylor1978}
who surveyed around 4000 samples from 100 species across different
kingdoms, it has been established that all species tend to spatially
aggregate. 

There are many causes for aggregation and determining these factors
is the fundamental subject of ecological and evolutionary theories
and of our understanding of the natural world. One cause that is often
disregarded is  encoded in the very nature of life: individuals appear
by birth close to their parents, but can die anywhere. Therefore,
each birth event enriches the spatial pair correlation function at
short distances, while a death event depletes all distances. Of course,
individuals (or their seed for plants) move randomly in space and
the diffusion phenomena tends to counteract and dilute the effect
of correlation created at short distance by birth. It can be shown
however \cite{Young2001,Houchmandzadeh2002,Houchmandzadeh2003,Houchmandzadeh2009a}that
diffusion is not enough to efficiently dilute the correlation creation
at short distances for \emph{spatial dimensions} $d\le3$ \emph{and}
large system size extension. This phenomena is called neutral clustering
(see \cite{Korolev2010a} for a review); it has been demonstrated
experimentally\cite{Houchmandzadeh2008} and the main concept has
been applied to other systems such as neutrons in nuclear reactors
\cite{Dumonteil2014,Houchmandzadeh2015c} and evolution of bimolecular
networks\cite{Stein2011}.

The same arguments can be applied to genome space. Consider individuals
characterized by the sequence of their genome, of length $L$. Each
duplication event can give rise to a new individual due to mutations
at one position in the sequence. The genetic distance $k$ between
two individuals being defined as the number of differences between
their sequence (the hamming distance), we see that birth events again
tend to enrich correlations at short distances, under the assumptions
that all mutations are \emph{neutral}, \emph{i.e.,} don't affect the
fitness of the individual. There is however a marked difference between
diffusion in real space due to random movements and diffusion in genotype
space due to mutations. The former happens in a low dimensional space
($d\le3$) and large extension; the latter occurs in high dimensional
space ($d=L$) but small extension (the number of values such as A,T,C,G
that a position along the sequence can take). Therefore, it is not
clear \emph{a priori} which effect (neutral aggregation or diffusion)
is dominant. The purpose of this article is to weight these factors
precisely in genotype space. 

In the evolutionary field, neutral models were first proposed by Kimura
\cite{Kimura1985}as the main driving force in evolution. In the ecology
field, Hubbell \cite{Hubbel2001} used a special version of the neutral
model (UNTB), called the infinite allele model \cite{KIMURA1964},
to explain the pattern of biodiversity in nature in terms of neutral
mutations . Both theories are passionately debated in the literature
(see \cite{Azaele2016} for a review of UNTB, \cite{Linquist2015}
for a critical review of both theories and their interconnections,
or \cite{Rosindell2011} defending the importance of neutral theories
in ecology). 

The advent of modern gene sequencing tools has opened the possibility
for more stringent tests of the neutral hypothesis in ecological communities,
combining the standard measurement of abundances of species with the
histogram of genetic distances between species. Jeraldo et al.\cite{Jeraldo2012}
for example measured the genetic distance between OTUs of six gastrointestinal
microbiomes of different mammals and found that the histogram of distances
is sharply peaked toward short distances. They thus concluded that
neutral processes play a negligible role in these communities and
selection is the dominant force. This argument of Jeraldo et al. is
however problematic, as they equate \emph{neutral} with ``drawn at
random'', ignoring the importance of neutral forces discussed above.
This point was raised by D'Andrea and Ostling\cite{DAndrea2016} who
developed a simple, interaction free model to demonstrate that neutral
causes for this model will lead to genetic distance histograms that
are peaked at short distances. 

We will show here, by exactly computing the distance pair correlation
function (normalized histogram) under the neutral hypothesis, that
the picture is more nuanced and depends crucially on the mutation
number, \emph{i.e. }the product of mutation probability and the population
size. We show that in well mixed populations, for small mutation number,
neutral aggregation is the dominant force and the distribution of
distances is nearly geometric, showing a sharp peak at the origin.
In this limit, most individuals are very close to each other in genotype
space. For large mutation numbers however, the distribution of distances
is more binomial-like and shows a peak at $\sim L/2$, as was assumed
by Jeraldo et al. (figure \ref{fig:Individual-based-stochastic}).

A further complication is that real populations are geographically
distributed and spatial migration also plays an important role. Histogram
measurements have thus to take into account this factor in order to
weight the effects of neutral factors. 

The aim of this article is to compute precisely the genetic pair correlation
function under the neutral hypothesis, using an individual based model.
This quantity is of prime importance in population genetics\cite{Watterson1975}
and is nowadays investigated mainly by coalescent theory\cite{Hein2005}.
However, application of coalescent theory to \emph{finite} sequences
and to geographically dispersed populations is rather difficult. In
contrast, the method we develop below uses only straightforward mathematical
tools and the results are derived by simple means. The computations
presented below are confirmed by individual based numerical simulations. 

This article is organized as follow: in the next section, we define
the model, the pair correlation function $u_{k}$ and summarize the
main findings of the article. The next three sections are devoted
to the mathematical derivations of these results: in section \ref{sec:Infinite-sequence-model},
we use the well known ``infinite site'' model to illustrate how
the equations governing $u_{k}$ can be obtained directly from the
model and be solved. Section \ref{sec:Finite-sequence-length.} generalizes
this approach to finite sequences when back mutations are explicitly
taken into account and the $u_{k}$ are found by their probability
generating function. This section constitutes the heart of this article.
Section \ref{sec:Spatially-distributed-population} generalizes this
approach to take into account the geographic extension of populations
and the influence of migrations. The final section is devoted to placing
this work in perspective and to the concluding remarks.

These computations should constitute a useful tool in assessing the
importance of neutral phenomena in ecological communities in general,
where results analogous to that obtained by Jeraldo et al. can be
evaluated.

\section{Model definition and main results.}

In this section, we define the model and summarize the main results
of the manuscripts. The mathematical derivations of these results
are given in the following sections. 

The two fundamental individual based models of population genetics
are the non-overlapping generations model of Wright-Fisher (WF)\cite{Fisher1999}
and the continuous time model of Moran\cite{Moran1962}. These models
are equivalent and obey the same diffusion equation in the large population
limit\cite{Houchmandzadeh2010}. Their use is dictated by the respective
ease by which relevant quantities are obtained. For the present computations,
we use the WF model which allows for the straightforward computation
of the pair correlation function and its extension to the spatial
case. 

\begin{figure}
\begin{centering}
\includegraphics[width=0.65\columnwidth]{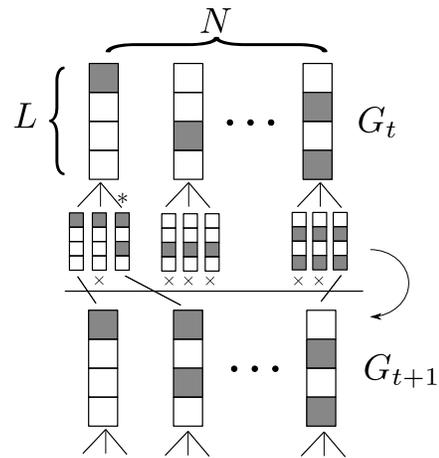}
\par\end{centering}
\caption{Neutral WF scheme for a well mixed population. The habitat contains
$N$ haploid individuals with various alleles of a gene of length
$L$. At generation $G_{t}$, each individual produces $q$ progeny,
some of whom may contain a mutation with respect to their parent (marked
by a star). $N$ individuals are selected among the $Nq$ progeny
to constitute the $G_{t+1}$ generation.\label{fig:NeutraWF}}

\end{figure}

Consider a habitat containing $N$ haploid individuals at each generation
(figure \ref{fig:NeutraWF}). The individuals are characterized by
the sequence of their genome of length $L$. Without loss of generality,
we will assume that each base can take only two values, $0$ and $1$:
When there are $2^{n}$ possible values for each base (such as $A,T,C,G$
where $n=2$ and $2^{n}=4$), we can map the problem to a binary system
with sequence length $nL$. 

At each generation, each individual produces $q>1$ progeny (the model
can be trivially generalized to a random number of progeny). The progeny
are then sampled at random to constitute the $N$ individuals of the
next generation (figure \ref{fig:NeutraWF}). We assume the system
to be \emph{neutral}: all individuals have equal fitness, and each
progeny, regardless of its genome, has the same probability of getting
to the next generation. 

Each progeny can differ from its parent because of mutations. In the
following, we will assume the mutation probability per base $\nu$
to be small and neglect the probability of having two mutations on
the same progeny. This limitation can be relaxed if needed. We further
assume $\nu$ to be the same for all bases of the genome sequence,
regardless of their position in the sequence and the value of neighboring
sites. This mutation model, known as K80, was first proposed by Kimura\cite{Kimura1980}
and various refinements of this model are widely used in the field
of molecular evolution, for example to deduce phylogenetic trees from
sequence data\cite{Yang2006}. 

\begin{figure}
\begin{centering}
\includegraphics[width=0.65\columnwidth]{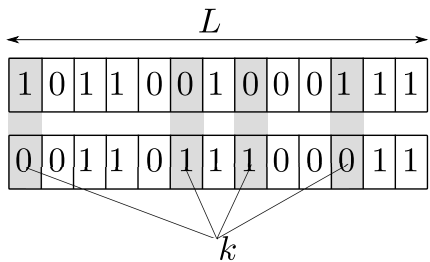}
\par\end{centering}
\caption{Two alleles of a gene of length $L$, differing in $k$ of their bases
(grayed area). A mutation in the grayed area will decrease the difference,
while a mutation in the white area will increase it. \label{fig:Two-alleles}}
\end{figure}

The \emph{genetic distance} between two individuals is defined as
the number of bases in which they differ (hamming distance). For example,
the two individuals shown in figure \ref{fig:Two-alleles} have a
hamming distance of $k=4$. 

The quantity we compute in the following sections is $u_{k}$, \emph{i.e.,}
the probability that, for large times, two individuals drawn at random
are at distance $k$ from each other. The computations are done in
the limit of small mutation probability per sequence $\lambda=L\nu\ll1$.
For a gene of length $L=1000$ base pair and mutation probability
per base of $\nu=10^{-10}$, this is a reasonable hypothesis. On the
other hand, for viruses with high mutation probabilities such as $\nu=10^{-4}$,
this approximation is more problematic and the model has to be extended
to take into account higher order perturbations. 

If the distribution $u_{k}$ is peaked at $k=0$ and is a fast decreasing
function of $k$, the system is clustered: there is one dominant sequence
in the population, and most other sequences are at short distances
from the dominant one. On the other hand, if sequences were distributed
totally randomly, we would expect $u_{k}$ to have a binomial distribution
\begin{equation}
u_{k}=2^{-L}\left(\begin{array}{c}
L\\
k
\end{array}\right)\label{eq:binomial0}
\end{equation}
and display a peak at $k=L/2$. 

We find that the actual distribution $u_{k}$ depends only on two
parameters: the sequence length $L$ and the mutation number \emph{per
base}
\begin{equation}
\Omega=2\nu N\label{eq:Theta-define0}
\end{equation}
The mutation number combines in a single number the contribution of
mutations and the population size. We find that for small mutation
numbers $\Omega<1/L$, the population is clustered, $u_{k}$ is peaked
at $k=0$ and decreases nearly exponentially with $k$. On the other
hand, for $\Omega>1$, the distribution of distances $u_{k}$ tends
toward a binomial one, showing peaks at $k>1$ (Figure \ref{fig:Individual-based-stochastic}). 

To measure experimentally $u_{k}$ necessitates a large quantity of
data. More robust quantities are the moments of the distribution such
as the mean distance $\left\langle k\right\rangle =\sum ku_{k}$ and
the variance $V=\sum k^{2}u_{k}-\left\langle k\right\rangle ^{2}$.
The moments of the distribution contain the same information as the
distribution itself. The results for the moments are particularly
simple. For example, the mean is given by 
\begin{equation}
\frac{\left\langle k\right\rangle }{L}\approx\frac{\Omega}{1+2\Omega}\label{eq:mean0}
\end{equation}
and all the other moments are deduced from the mean by simple recurrence
( equation \ref{eq:kn:recur}). The probability distribution $u_{k}$
can be determined from these moments (equation \ref{eq:ul:detail}). 

We see again from expression (\ref{eq:mean0}) that for $\Omega<1/L$,
$\left\langle k\right\rangle <1$ and that most sequences are located
at very short distances from the dominant one. On the other hand,
for $\Omega\gg1$, $\left\langle k\right\rangle \approx L/2$, a result
which is expected from the binomial distribution of $u_{k}$ (relation
\ref{eq:binomial0}) and which denotes a random distribution of sequences. 

We observe that the distribution depends crucially on the size of
the population $N$. For large populations, the hypothesis of well
mixed populations ceases to be valid. We must explicitly take into
account physical space and migrations in order to determine the effective
size of the population. 

\begin{figure}
\begin{centering}
\includegraphics[width=0.65\columnwidth]{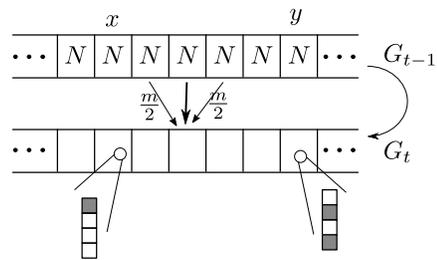}
\par\end{centering}
\caption{Scheme of the spatial WF model for finite sequence length. Space is
divided into patches of $N$ individuals. In each patch, the WF stochastic
process described in fig. \ref{fig:NeutraWF} takes place, with the
additional condition that a progeny can come, with probability $m/2$
from one of the neighboring patches. $u(x,y;k;t)$ is the probability
that two individuals drawn at random in patches $x$ and $y$ at time
$t$ are at genetic distance $k$. \label{fig:Scheme-spatial}}
\end{figure}

For this purpose, we resort to the classical stepping stone model
(see \cite{Korolev2010a} for a review), dividing the space into demes
(or patches) of $N$ individuals, and denote by $m$ the migration
probability, \emph{i.e.} the probability that an individual in a deme
at generation $t+1$ has its parent in a neighboring deme at generation
$t$ (figure \ref{fig:Scheme-spatial}). The \emph{migration number}
is defined as $M=Nm$ and combines into a single number the size of
the deme and the migration probability. 

Let $y$ be the number of patches between two demes. We compute $u(k,y)$,
the probability that two individuals drawn at random in two demes
at physical distance $y$ are at genetic distance $k$. We restrict
the computation to the mean hamming distance 
\[
\left\langle k(y)\right\rangle =\sum_{k=0}^{L}ku(k,y)
\]
We show that the mean hamming distance in the same patch is similar
to expression (\ref{eq:mean0}) 
\begin{equation}
\frac{\left\langle k(0)\right\rangle }{L}=\frac{\Omega_{e}}{1+2\Omega_{e}}\label{eq:ky0}
\end{equation}
Where the effective mutation number $\Omega_{e}$ is defined as 
\[
\Omega_{e}=\sqrt{\Omega^{2}+2\Omega M}
\]
The amount of neutral clustering depends crucially on competition
between migration and mutations. Moreover, $\left\langle k(y)\right\rangle $
tends exponentially in $y$ toward $L/2$ (relation \ref{eq:mean:ky:solved}). 

In the following sections, we detail the mathematical derivations
of the above results and provide a rough estimation of the expected
clustering for gut bacteria. 

\section{Infinite sequence model.\label{sec:Infinite-sequence-model}}

Consider a well mixed population of $N$ individuals, where each individual
is characterized by the sequence of its genome of length $L$; the
population follows a Wright-Fisher competition (figure \ref{fig:NeutraWF}).
At each reproduction event, a base (a position along the sequence)
can be mutated with probability $\nu$. We define the mutation probability
per \emph{sequence} as 
\[
\lambda=L\nu
\]
and we suppose $\lambda\ll1$ in this article (see below for the limit
of validity of this hypothesis).

We wish to compute the probability $u_{k}(t+1)$ that two individuals
drawn at random at generation $G_{t+1}$ are at (genetic) distance
$k$, \emph{i.e.} they differ in exactly $k$ bases. 

Before doing the full computation, let us consider the classical ``infinite
site'' model\cite{Watterson1975,Tajima1996}, where the main mathematical
concepts are straightforward to introduce. The infinite site model
assumes that each mutation gives rise to a new individual that does
not already exist in the habitat; this consists in assuming $L\rightarrow\infty$
while $\lambda$ remains finite. 

We denote 

\begin{equation}
a=\frac{q-1}{Nq-1}\label{eq:sameparent}
\end{equation}
as the probability that two individuals drawn at random are from the
same parent, where $q$ is the number of progeny of each individual.
Note that for $q\gg1$, $a\approx1/N$. 

The probability that two individuals are at distance $k=0$ from each
other is :
\begin{enumerate}
\item The two individuals are from the same parent (probability $a$) AND
no mutation has taken place in them (probability $(1-\lambda)^{2}$)
OR 
\item The two individuals are from different parents which themselves are
genetically identical (probability $(1-a)u_{0}(t)$ ) AND no mutation
has taken place in them (probability $(1-\lambda)^{2}$).
\end{enumerate}
In other words, 

\begin{equation}
u_{0}(t+1)=(1-\lambda)^{2}\left[a+(1-a)u_{0}(t)\right]\label{eq:inf-u0}
\end{equation}
Following the same line of argument (see appendix \ref{sec:The-infinite-allele}),
we can write generally the linear system:
\begin{eqnarray}
u_{0}(t+1) & = & (1-a)Bu_{0}(t)+b_{0}\label{eq:infb-u0}\\
u_{1}(t+1) & = & (1-a)\left(Au_{0}(t)+Bu_{1}(t)\right)+b_{1}\label{eq:infb-u1}\\
u_{k}(t+1) & = & (1-a)\left(Au_{k-1}(t)+Bu_{k}(t)\right)\,\,\,\,\,\,(k>1)\label{eq:infb-uk}
\end{eqnarray}
where $A=2\lambda(1-\lambda)$, $B=(1-\lambda)^{2}$, $b_{0}=(1-\lambda)^{2}a$
and $b_{1}=2\lambda(1-\lambda)a$. Note that the probability that
two individuals from the same parent are at distance $k=2$ is $a\lambda^{2}$
and therefore it has been neglected in relation (\ref{eq:infb-uk}).
It should also be noted that because we suppose $\lambda\ll1$, $A/B=b_{1}/b_{0}\sim2\lambda\ll1$.
The coefficient $A$ captures mutation events that \emph{increase}
the genetic distances, while $B$ relates to events that maintain
the genetic distance. 

We define 
\[
\text{\ensuremath{\Phi}}=\frac{(1-a)A}{1-(1-a)B}\approx\frac{2\lambda}{a+2\lambda}=\frac{\Theta}{1+\Theta}\,\,\,\in[0,1)
\]
where 
\begin{equation}
\Theta=2\lambda/a\approx2\lambda N\label{eq:def:Theta}
\end{equation}
is the \emph{per sequence} mutation number. The stationary solution
$u_{k}$ for large times of relations (\ref{eq:infb-u0}-\ref{eq:infb-uk}),
which is a one term recurrence relation, is:
\begin{eqnarray}
u_{k} & = & \Phi^{k}(u_{0}+\frac{a}{1-a})\,\,\,(k\ge1)\label{eq:infal-sol-1}\\
 & \approx & \Phi^{k}(u_{0}+\frac{1}{N})\label{eq:infal-sol-2}
\end{eqnarray}
where 
\begin{equation}
u_{0}=\frac{b_{0}}{1-(1-a)B}\approx\frac{a}{a+2\lambda}=\frac{1}{1+\Theta}\label{eq:infal-sol-u0}
\end{equation}
Note that for the infinite allele model, the autocorrelation function
is always peaked at $k=0$ (or at $k=1$ for $\Theta\gg1$). In the
low mutation regime ($\Theta\ll1$), nearly all individuals are identical
($u_{0}\approx1-\Theta$), and $u_{k}$ drops sharply as a function
of $k$ ($\Phi\ll1$). On the other hand, in the high mutation regime
$\Theta\gg1$, the distribution is nearly flat for a wide range of
$k$, and many different genomes coexist. The mean genetic distance
for this model is : 
\[
\left\langle k\right\rangle =\sum_{k=1}^{\infty}ku_{k}\approx\Theta(1+a\Theta)
\]

In the above computation, we have assumed $\lambda\ll1$ and therefore
have neglected terms in $\lambda^{2}$ . We can evaluate the accuracy
of this approximation by estimating $S(t)=\sum_{k}u_{k}(t)$ and its
deviation from unity. Summing the lines of relations (\ref{eq:infb-u0}-\ref{eq:infb-uk}),
we have 
\begin{eqnarray*}
S(t+1) & = & (1-a)(A+B)S(t)+(b_{0}+b_{1})\\
 & = & (1-\lambda^{2})\left[(1-a)S(t)+a\right]
\end{eqnarray*}
and for large times, the stationary value of $S$ is 
\[
S\approx\frac{a}{a+\lambda^{2}}\approx1-\frac{\lambda^{2}}{a}\approx1-N\lambda^{2}
\]
Therefore the approximation we are making is valid when $N\lambda^{2}\ll1$.
For gene sequences of length $L\sim10^{3}$ and mutation probability
per base $\nu\approx10^{-9}$, the limit of validity of this computation
is $N\ll10^{12}$. The approximation is therefore valid for large
populations, especially when we consider spatially distributed populations
as in section \ref{sec:Spatially-distributed-population}. 

\section{Finite sequence length.\label{sec:Finite-sequence-length.}}

We now investigate the case of finite sequence length $L$, which
is the main point of this article. Note that if sequences were generated
randomly with equal probabilities for $0$ and 1 values of the bases,
the probabilities $u_{k}$ would follow the binomial distribution
\begin{equation}
u_{k}=2^{-L}\left(\begin{array}{c}
L\\
k
\end{array}\right)\label{eq:binomial}
\end{equation}
where the mean distance is $\left\langle k\right\rangle =L/2$ and
the variance $V=L/4$. This is the implicit hypothesis used by Jeraldo
et al. to exclude neutral processes. However sequences are not generated
randomly, but are inherited from parents with a small probability
for mutations. This fact radically changes the expected distribution
of distances as we show below.

\subsection{The evolution equation.}

In the infinite site model considered in the preceding section, we
captured the basic equations (\ref{eq:infb-u0}-\ref{eq:infb-uk})
in terms of events that increase the genetic distance (coefficient
$A$) and events that maintain the genetic distance (coefficient $B$).
For finite sequences, the picture is more complicated. First, mutations
can also \emph{decrease} genetic distances if they appear in bases
which are already different between two sequences (back mutations);
we would capture these events in a coefficient $C$. Second, the coefficients
are not constant but depend on the number of bases in which two individuals
already differ. 

Consider two sequences of length $L$ that differ in $k$ bases (figure
\ref{fig:Two-alleles}). The probability that an event increases their
difference by one unit is that of one mutation ( probability $2\lambda(1-\lambda)$
) occurring in one of their identical bases (probability $(L-k)/L$
) : 
\begin{equation}
A_{k}=2\lambda(1-\lambda)(1-\frac{k}{L})=2\lambda(1-\frac{k}{L})+O(\lambda^{2})\label{eq:Ak}
\end{equation}
By the same token, the probability of decreasing their distance by
one unit is 
\begin{equation}
C_{k}=2\lambda(1-\lambda)\frac{k}{L}=2\lambda\frac{k}{L}+O(\lambda^{2})\label{eq:Ck}
\end{equation}
The probability that their distance is conserved is 
\begin{equation}
B_{k}=(1-\lambda)^{2}+2\lambda^{2}\frac{k(L-k)}{L^{2}}=1-2\lambda+O(\lambda^{2})\label{eq:Bk}
\end{equation}
and to the first order in $\lambda$, $A_{k}+B_{k}+C_{k}=1$. Now,
following the same line of argument as in the preceding section, the
probability $u_{k}(t+1)$ of finding two individuals at distance $k$
at time $t+1$ is 
\begin{eqnarray}
u_{0}(t+1) & = & (1-a)\left(B_{0}u_{0}(t)+C_{1}u_{1}(t)\right)+b_{0}\label{eq:fin-u0}\\
u_{1}(t+1) & = & (1-a)\left(A_{0}u_{0}(t)+B_{1}u_{1}(t)+C_{2}u_{2}(t)\right)+b_{1}\label{eq:fin-u1}\\
u_{k}(t+1) & = & (1-a)\left(A_{k-1}u_{k-1}+B_{k}u_{k}+C_{k+1}u_{k+1}\right)\label{eq:fin-uk}\\
u_{L}(t+1) & = & (1-a)(A_{L-1}u_{L-1}+B_{L}u_{L})\label{eq:fin-uL}
\end{eqnarray}
where $b_{0}$ and $b_{1}$ are defined as before and $1<k<L$ in
equation (\ref{eq:fin-uk}). Note that obviously, $u_{k}=0$ for $k>L$.
The above set of linear equations can be written in vectorial notation
as
\begin{equation}
\left|u(t+1)\right>=(1-a)Q\left|u(t)\right>+\left|b\right>\label{eq:vectorial}
\end{equation}
where $\left|u(t)\right>=(u_{0}(t),u_{1}(t),...,u_{L}(t))^{T}$,
\begin{equation}
Q=\left(\begin{array}{cccccc}
B_{0} & C_{1} & 0 & 0 & \cdots & 0\\
A_{0} & B_{1} & C_{2} & 0 & \cdots & 0\\
0 & A_{1} & B_{2} & C_{3} & \ddots & \vdots\\
0 & 0 & \ddots & \ddots & \ddots & 0\\
\vdots &  & \ddots & \ddots & \ddots & C_{L}\\
0 &  & \cdots & 0 & A_{L-1} & B_{L}
\end{array}\right)\label{eq:Q}
\end{equation}
and $\left|b\right>=(b_{0},b_{1},0,\ldots,0)^{T}$. The stationary
probabilities $\left|u\right>$ for large times are obtained from
\begin{equation}
\left(I-(1-a)Q\right)\left|u\right>=\left|b\right>\label{eq:u:stationary}
\end{equation}
where $I$ is the identity matrix. The above relation is a $(L+1)\times(L+1)$
linear system that can be solved numerically for practical purposes.
We are of course interested in its analytical solution. 

\subsection{The probability generating function.}

The non-homogeneous linear system (\ref{eq:vectorial}) is a two term
recurrence relation where the coefficients are not constant. The stationary
solution $u_{k}$ for large time however is surprisingly simple if
we use the probability generating function (PGF)
\begin{equation}
\phi(z)=\sum_{k=0}^{L}u_{k}z^{k}\label{eq:phidef}
\end{equation}
The PGF contains the most complete information on the system; the
probabilities and their moments are obtained from the derivatives
of $\phi$ at either $z=0$ or $z=1$. Let 
\begin{equation}
\left\langle k_{(n)}\right\rangle =\left\langle k(k-1)...(k-n+1)\right\rangle \label{eq:def:kn}
\end{equation}
be the factorial moment of order $n$. For example, the usual mean
distance is $\left\langle k\right\rangle =\left\langle k_{(1)}\right\rangle $
and the variance is $V=\left\langle k_{(2)}\right\rangle +\left\langle k\right\rangle -\left\langle k\right\rangle ^{2}$
and so on. The factorial moments are given by 
\begin{equation}
\left\langle k_{(n)}\right\rangle =\left.\frac{d^{n}\phi}{dz^{n}}\right|_{z=1}\label{eq:factorialmomentdef}
\end{equation}
The moments, specially the lower ones, are the most robust quantities
that we can estimate from real data analysis. Below, we shall also
use the normalized factorial moments $\mu_{n}=\left\langle k_{(n)}\right\rangle /n!$
. On the other hand, the probabilities are 
\begin{equation}
u_{k}=\frac{1}{k!}\left.\frac{d^{k}\phi}{dz^{k}}\right|_{z=0}\label{eq:probadef}
\end{equation}
although experimentally, their estimations necessitates much more
data than what is needed for (lower) moment estimation.

It can be shown that the PGF obeys a simple first order differential
equation (see appendix \ref{sec:The-PGF-equation.}): 
\begin{equation}
(1-z^{2})\phi'+\left(Lz-(1+\frac{1}{\Theta})L\right)\phi=-aL(z-1)-\frac{L}{\Theta}\label{eq:phi}
\end{equation}
The term $aL\approx L/N$ weights the relative importance of the sequence
length compared to the population size. The above differential equation
can be exactly solved in terms of the hypergeometric function. However,
as we are interested only in the probabilities $u_{k}$ and their
moments, we don't even need to solve equation (\ref{eq:phi}) and
we can extract all the moments from simple arguments as discussed
below.

Before the full discussion, note that $N\rightarrow\infty$ implies
$\Theta\rightarrow\infty$ and $a\rightarrow0$. In this high mutation
number regime, equation (\ref{eq:phi}) becomes 
\[
(1-z^{2})\phi'+L(z-1)\phi=0
\]
 with the obvious solution 
\[
\phi(z)=2^{-L}(z+1)^{L}
\]
satisfying the initial condition $\phi(1)=1$. This is the PGF for
the binomial distribution, which is expected if sequences were drawn
at random. Therefore, in the very high mutation number regime ($\Theta\gg L$),
the distribution of distances indeed becomes binomial. 

On the other hand, the infinite sites results can be recovered from
equation (\ref{eq:phi}) by letting $L\rightarrow\infty$. In this
limit, the $\phi'$ term becomes negligible in equation (\ref{eq:phi})
and the PGF is simply 
\begin{equation}
\phi(z)=\frac{1-a\Theta+a\Theta z}{1+\Theta-\Theta z}\approx\frac{1+a\Theta z}{1+\Theta-\Theta z}\label{eq:pgf:L:large}
\end{equation}
where $a\Theta=2\lambda$ has been neglected compared to one, as we
suppose $\lambda\ll1$. Relation (\ref{eq:pgf:L:large}) is the PGF
of the probabilities $u_{k}$ computed for infinite sites model in
the preceding section (equation \ref{eq:infal-sol-2}). This expression
was first computed by Watterson\cite{Watterson1975} for the infinite
allele model. 

\subsection{Finding the moments and probabilities.}

Let us first consider the point $z=1$ in equation (\ref{eq:phi}).
The first term vanishes at this point and we have trivially $\phi(1)=1$
which just states that the sum of the probabilities is unity: 
\[
\phi(1)=\sum_{k=0}^{L}u_{k}=1
\]
We see here that computing $\phi(1)$ does not require a knowledge
of $\phi'(1)$. This is a general feature of the PGF equation (\ref{eq:phi}):
$\phi^{(n)}(1)$ does not depend on the $\phi^{(n+1)}(1)$ and we
can deduce all the moments from a hierarchical structure. To see this,
we differentiate equation (\ref{eq:phi}) in respect to $z$:
\begin{equation}
(1-z^{2})\phi"+\left((L-2)z-(1+\frac{1}{\Theta})L\right)\phi'+L\phi=-aL\label{eq:phiprime}
\end{equation}
As before, the higher order term vanishes at $z=1$ and therefore
the mean distance between individuals is 
\begin{equation}
\left\langle k\right\rangle =\phi'(1)=\frac{L\Theta(a+1)}{L+2\Theta}\approx\frac{L\Theta}{L+2\Theta}\label{eq:mean}
\end{equation}
we see here that for mutation numbers $\Theta<1$, the mean distance
$\left\langle k\right\rangle <1$ and most individuals are clustered
very close to each other in genetic space. Only for high mutation
numbers $\Theta=O(L)$, does the mean distance between individuals
become appreciable. For example, $\Theta=L/2$ results in $\left\langle k\right\rangle =L/4$.
For very large mutation numbers $\Theta\gg L$ we reach $\left\langle k\right\rangle =L/2$,
as in the binomial distribution. 

Obtaining higher moments follows the same procedure. Applying $d/dz$
to equation (\ref{eq:phiprime}) and computing $\phi''(1)$, we find
\[
\left\langle k_{(2)}\right\rangle =\frac{2L(L-1)\Theta^{2}}{(L+2\Theta)(L+4\Theta)}
\]
For the low mutation regime $\Theta\lesssim1$ and long sequences
$L\gg1$, $\left\langle k_{(2)}\right\rangle \approx2\Theta^{2}$
and the distance distribution is indeed sharply centered around the
origin. For the high mutation regime $\Theta\gg L$, we find $\left\langle k_{(2)}\right\rangle =L(L-1)/4$,
as expected from the binomial distribution. 

Successive differentiation allows us to obtain all the moments. For
$n\ge2$
\begin{equation}
\left\langle k_{(n)}\right\rangle =\frac{n(L-n+1)\Theta}{L+2\Theta n}\left\langle k_{(n-1)}\right\rangle \label{eq:kn:recur}
\end{equation}
which leads to 
\begin{eqnarray}
\mu_{n}=\frac{1}{n!}\left\langle k_{(n)}\right\rangle  & = & 2^{-n}\frac{L(L-1)...(L-n+1)}{(\gamma+1)(\gamma+2)...(\gamma+n)}\label{eq:kn:expand}\\
 & = & 2^{-n}\frac{(L)_{(n)}}{(\gamma+n)_{(n)}}\,\,\,\,\,\,\,\,\,\,\,\,(n\ge1)\label{eq:kn:poch}
\end{eqnarray}
where $\gamma=L/(2\Theta)$ and $(r)_{(n)}$ is the descending Pochhammer
symbol
\[
(r)_{(n)}=r(r-1)...(r-n+1)=\frac{\Gamma(r+1)}{\Gamma(r+1-n)}
\]
The expression (\ref{eq:kn:expand},\ref{eq:kn:poch}) for $\left\langle k_{(n)}\right\rangle $
should be corrected by the factor $(1+a)$ for very small populations,
as in relation (\ref{eq:mean}). 

\begin{figure}
\begin{centering}
\includegraphics[width=0.95\columnwidth]{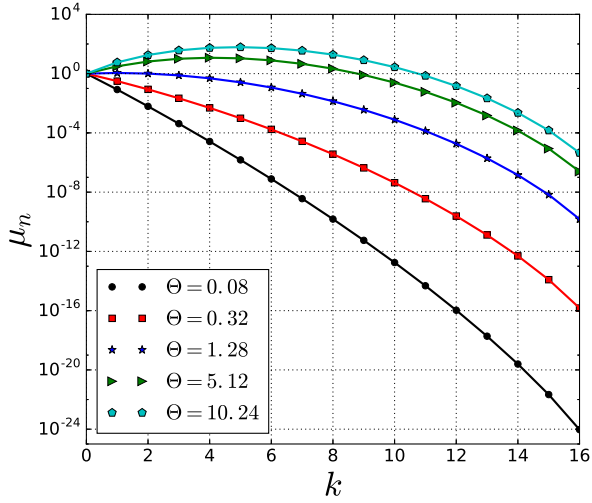}
\par\end{centering}
\caption{(Color online) The normalized factorial moments $\mu_{n}$ as a function
of $n$ for different values of $\Theta$. $L=16$, $\lambda=2.5\times10^{-3}$
and $a=1/N=2\lambda/\Theta$. Solid lines represent theoretical values
given by relation (\ref{eq:kn:poch}). Symbols are obtained by solving
the linear system (\ref{eq:u:stationary}) for stationary probabilities
and then computing their factorial moments from equation (\ref{eq:def:kn}).
\label{fig:norm:moment} }
\end{figure}

Figure \ref{fig:norm:moment} shows the results for normalized factorial
moments $\mu_{n}$ obtained from relation (\ref{eq:kn:poch}) and
their excellent agreement with the moments obtained by numerical resolution
of the linear set of equations (\ref{eq:u:stationary}).

\begin{figure}
\begin{centering}
\includegraphics[width=0.8\columnwidth]{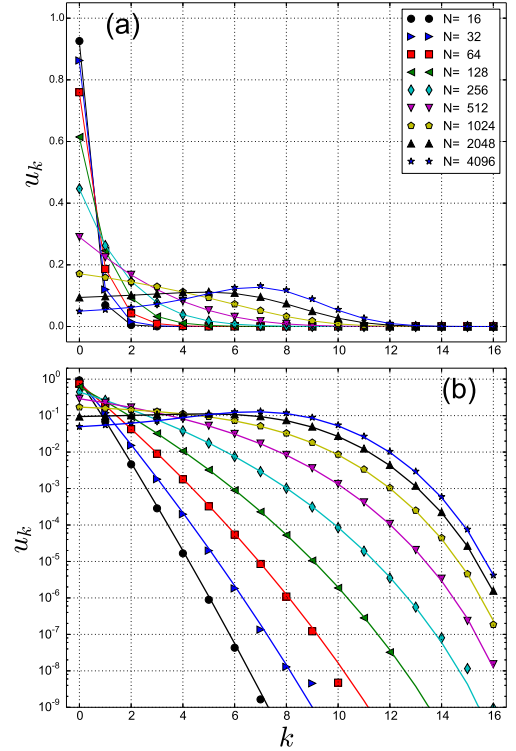}
\par\end{centering}
\caption{(Color online) Individual based stochastic simulation (symbols) of
the finite sequence neutral WF model and its comparison with the theoretical
predictions (solid lines). Figures show the probabilities $u_{k}$
as a function of $k$ for different values of $N$ (and hence $\Theta$)
in (a) linear, (b) logarithmic representations. $L=16,$ $\lambda=2.5\times10^{-3}$,
$N\in[16,4096],$ $\Theta\in[0.08,20.48]$. Simulation values are
obtained by generating between $10^{6}$(for large $N$) to $10^{8}$(for
smaller $N$) paths for $10N$ generations (see appendix \ref{sec:Numerical-simulations}).
\label{fig:Individual-based-stochastic}}

\end{figure}

To find the probabilities $u_{k}$, we can rearrange the PGF function
\[
\phi(z)=\sum_{n=0}^{L}u_{n}z^{n}=\sum_{n=0}^{L}\mu_{n}(z-1)^{n}
\]
Expanding $(z-1)^{n}$ and identifying the corresponding powers of
$z$ in both sums, we find
\begin{equation}
u_{\ell}=\sum_{n=\ell}^{L}(-1)^{n-\ell}\mu_{n}\left(\begin{array}{c}
n\\
\ell
\end{array}\right)\label{eq:ul:detail}
\end{equation}
Defining the matrix $C$ such that $(C)_{\ell}^{n}=(-1)^{n-\ell}\left(\begin{array}{c}
n\\
\ell
\end{array}\right)$, the solution in vectorial notation is 
\begin{equation}
\left|u\right>=C\left|\mu\right>\label{eq:ul:vectorial}
\end{equation}
Figure \ref{fig:Individual-based-stochastic} shows the excellent
agreement between the above theoretical results and individual based
stochastic simulations of the neutral model (see appendix \ref{sec:Numerical-simulations}).

It can be observed that as $\Theta$ increases , $u_{k}$ transforms
from a sharply decreasing distribution with its peak at zero toward
a binomial distribution with the peak approaching $L/2$. The curves
cease to decrease monotonically as a function of the genetic distance
$k$ for $\Theta\approx L/2$. 

For small mutation numbers $\Theta\ll1$, $\mu_{n}$ given by expression
(\ref{eq:kn:poch}) can be approximated by 
\begin{equation}
\mu_{n}=(1+a)\left(\frac{\Theta}{L}\right)^{n}(L)_{(n)}\,\,\,\,n\ge1\label{eq:mun:approx}
\end{equation}
which decreases faster than exponentially. In this case, only the
first term contributes to the sum (\ref{eq:ul:detail}) and we have
\begin{eqnarray*}
u_{0} & = & 1-(1+a)\Theta\\
u_{k} & = & \mu_{k}\,\,\,\,\,(k\ge1)
\end{eqnarray*}
It has been checked numerically that this approximation of the probabilities
$u_{k}$ is very good for $\Theta<0.1$. 

To summarize this section, we have obtained the exact solution for
the moments $\mu_{n}$ (relation \ref{eq:kn:poch}) and probabilities
$u_{k}$ (relation \ref{eq:ul:vectorial}) of the finite sequence
neutral WF model; we have obtained various limiting expressions (large
and small $\Theta$, large $L$) and have shown the accuracy of these
expressions by comparing them with numerical simulations. The parameter
$\Omega$ that controls the behavior of the system is the ratio of
mutation number to sequence length 
\begin{equation}
\text{\ensuremath{\Omega}=}\frac{\Theta}{L}=2N\nu\label{eq:Omega:def}
\end{equation}
\emph{i.e., }the \emph{per base} mutation number. The expressions
for moments could have been derived by using this number instead.
For example, the relative mean distance is 
\begin{equation}
\frac{\left\langle k\right\rangle }{L}=\frac{\Omega}{1+2\Omega}\label{eq:mean:alt}
\end{equation}

For bacteria such as $E.Coli$ present at $10^{10}$/ml in the human
gut for example\cite{Baron1996} and overall mutation probability
at $\approx10^{-10}$ per \emph{base}\cite{Wielgoss2011}, the mutation
number is $\Theta\sim2L$ for a population contained in 1ml. In this
regime, we expect to find a large distribution of distances with a
peak around $L/2$, as was indeed supposed by Jeraldo et al.\cite{Jeraldo2012}.
On the other hand, the 1 ml volume choice is arbitrary, and if we
had considered a $1\mu$l volume instead, we would be in the low mutation
regime where neutral aggregation dominates. In order to determine
the effective mutation number $\Omega_{e}$ needed to test the neutral
hypothesis, we must explicitly take into account the dispersion of
individuals in the \emph{real} space. The next section deals with
this problem.

\section{Spatially distributed populations.\label{sec:Spatially-distributed-population}}

Consider a population distributed into patches, each site containing
$N$ individuals. In each site, the same WF stochastic process as
discussed above takes place, with the additional condition that a
progeny can descend from a parent in a neighboring site with probability
$m/2d$, where $d$ is the dimension of real space. To take space
into account, we need to compute $u(x,y;k;t)$, the probability of
finding two individuals drawn at random in sites $x$ and $y$ at
time $t$ at genetic distance $k$. 

\begin{figure}
\begin{centering}
\includegraphics[width=0.95\columnwidth]{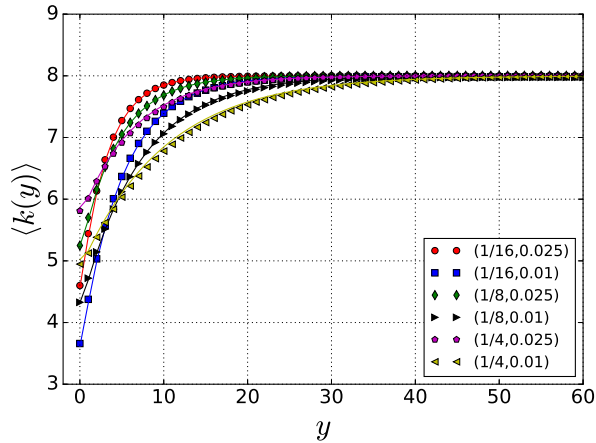}
\par\end{centering}
\caption{(Color online) Comparison between the theoretical result for $\left\langle k(y)\right\rangle $
(solid line) and stochastic simulations of the spatial, neutral, finite
sequence length of WF model (symbols) for various value of $(m,\lambda)$.
There are $Q=128$ one dimensional, circular sites with $N=32$ individual
per site and $L=16$. \label{fig:spatial}}

\end{figure}

Let us first consider a one dimensional real space $d=1$. We assume
translational invariance:
\[
u(x,x';k;t)=u(|x-x'|;k;t)
\]
and will compute $u(y;k;t)$ where $y$ is the absolute discrete distance
between two patches. The evolution equation (\ref{eq:vectorial})
for $d=0$ of the previous section was derived by the usual combination
of AND,OR, distinguishing the cases where two individuals descend
from the same parent or not. For the spatial case, we must also distinguish
the cases where the parent can be from the same patch or not. Again,
we collect $u(y;k;t)$ ($0\le k\le L$ ) into a vector 
\[
\left|u(y;t)\right>=u\left((y;0;t),u(y;1;t),...,u(y;L;t)\right)^{T}
\]
Following the same line of arguments as the preceding section, the
evolution equation (\ref{eq:vectorial}) is generalized for the spatial
case :
\begin{eqnarray}
\left|u(0;t)\right> & = & (1-2m)\left\{ (1-a)Q\left|u(0;t)\right>+\left|b\right>\right\} \label{eq:uyk:0}\\
 & + & 2mQ\left|u(1;t)\right>\nonumber \\
\left|u(1;t)\right> & = & m\left\{ (1-a)Q\left|u(0;t)\right>+\left|b\right>\right\} \label{eq:uyk:1}\\
 & + & (1-2m)Q\left|u(1;t)\right>+mQ\left|u(2;t)\right>\nonumber \\
\left|u(y;t)\right> & = & mQ\left|u(y-1;t)\right>+(1-2m)Q\left|u(y;t)\right>\label{eq:uyk:y}\\
 & + & mQ\left|u(y+1;t)\right>\,\,\,\,\,(y\ge2)\nonumber 
\end{eqnarray}
As before, the stationary solution of these equations for large times
can be obtained by using the spatial PGF function 

\begin{equation}
\phi(y;z)=\sum_{k=0}^{L}u(y;k)z^{k}\label{eq:PGFspatial:def}
\end{equation}
By its very definition, $\phi(y;1)=1$ $\forall y$. An evolution
equation can be obtained for the PGF which has the same property as
in the preceding section: $\left.\left(\partial^{n}/\partial_{z^{n}}\right)\phi(y,z)\right|_{z=1}$
depends only on the lower derivatives, and all the factorial moments
can be obtained from a hierarchical structure. For example, the mean
\[
\left\langle k(y)\right\rangle =\sum_{k=0}^{L}ku(y,k)=\left.\frac{\partial\phi(y,z)}{\partial z}\right|_{z=1}
\]
is shown to obey the equation 
\begin{eqnarray}
\left(-r+m\Delta\right)\left\langle k(y)\right\rangle  & = & -\frac{Lr}{2}\label{eq:mean:ky}\\
 & + & a\left\{ (1-2m)\delta_{y,0}+m\delta_{y,1}\right\} \left\langle k(0)\right\rangle \nonumber 
\end{eqnarray}
 where $\Delta$ is the discrete Laplacian operator 
\[
\Delta f(y)=f(y-1)-2f(y)+f(y+1)
\]
 and 
\[
r=\frac{2a\Theta}{L-2a\Theta}\approx4\frac{\lambda}{L}=4\nu
\]
Equation (\ref{eq:mean:ky}) is a non-homogeneous two terms recurrence
equation, which can easily be solved. The solution depends on the
per sequence mutation number $\Omega=2N\nu=\Theta/L$ and the migration
number 
\begin{equation}
M=Nm\label{eq:migration:number}
\end{equation}
and is written 
\begin{equation}
\frac{\left\langle k(y)\right\rangle }{L}=\frac{1}{2}-\left[\frac{1}{2}-(1-a)\frac{\left\langle k(0\right\rangle }{L}\right]s^{y}\,\,\,\,(y>0)\label{eq:mean:ky:solved}
\end{equation}
where $s$ is the less than unity solution of 
\begin{equation}
s^{2}-2(1+\frac{\Omega}{M})s+1=0\label{eq:ell}
\end{equation}
and
\begin{equation}
\frac{\left\langle k(0)\right\rangle }{L}=\frac{\Omega+M(1-s)}{1+2\Omega+2M(1-s)}\label{eq:mean:k0:solved}
\end{equation}
in the regime where $a\ll1$ and $m\ll1$ (see appendix \ref{sec:The-PGF-equation-space}
for the full expression). Defining the effective mutation number as
\begin{eqnarray}
\Omega_{e} & = & \Omega+M(1-s)\label{eq:Omega:effective}\\
 & = & \sqrt{\Omega^{2}+2\Omega M}
\end{eqnarray}
we see that the expression for $\left\langle k(0)/L\right\rangle $
is similar to the case of a well mixed population (equation \ref{eq:mean:alt})
where $\Omega$ has been replaced by $\Omega_{e}$. Figure \ref{fig:spatial}
shows the excellent agreement between the above theoretical expressions
and the results from individual based numerical solution of the spatial,
finite sequence length WF model.

We first note that for large separations $y\gg1$ between sites, $\left\langle k(y)\right\rangle \approx L/2$.
This result is expected, as far-distant sites evolve independently
of each other and we expect to recover the binomial distribution for
$u(y,k)$ when $y\gg y^{*}=-1/\log(s)$. 

On the other hand, $\left\langle k(0)\right\rangle $ is given by
a balance between mutation number $\Omega$ and migration number $M$,
and their relative amplitude with respect to unity. Even when $\Omega\ll1$,
$\left\langle k(0)\right\rangle $ can be substantially greater than
unity if the migration number is not too small. In particular, note
that $\Omega/M=2\nu/m$; when $\nu\ll m$, the effective mutation
number is 
\[
\Omega_{e}=\sqrt{\Omega^{2}+2\Omega M}\approx\sqrt{2\Omega M}=2N\sqrt{\nu m}
\]
and the typical distance under which the mean distance notably deviates
from $L/2$ is 
\[
y^{*}=-1/\log s\approx\sqrt{m/(4\nu)}
\]

Higher moments can be found by the same method if needed, although
the algebra becomes increasingly cumbersome. The extension to higher
dimensions is trivial and necessitates only the redefinition of the
discrete Laplacian operator $\Delta$ (see appendix \ref{sec:The-PGF-equation-space}). 

We can now try to make a rough estimation of the order of magnitude
of the neutral clustering for the bacterial communities studied by
Jeraldo et al. Let us suppose that individuals (or their seed) diffuse
in one generation according to a dispersal law which for simplicity
we take to be a Gaussian 
\[
p_{s}(x)=(2\pi\sigma^{2})^{-1/2}e^{-x^{2}/2\sigma^{2}}
\]
The dispersion length $\sigma$ determines both $N$ and $m$. We
can model the space as divided into patches of size $\ell=3\sigma$.
The exact size of the patch is not crucial, we have to choose it in
a way that insures that (i) there is negligible migration to the next
nearest neighbor and (ii) every progeny has a non-negligible probability
of descending from every parent inside the same patch. $\ell=3\sigma$
is a good compromise. For patches of this size, the migration probability
between patches is $m\approx0.27$ and the number of individual inside
each patch is $N=\ell c_{0}$, where $c_{0}$ is the bacterial concentration.
Note that the migration number $Nm$ is not sensitive to the choice
of $\ell$. 

We can, as an order of magnitude, choose 1 $\mu$m as the bacterial
size; in very dense bacterial concentrations, bacterial movements
are reduced and we may broadly estimate $\sigma$ as some 10 times
the bacterial size and therefore $N=30$. With this choice of $N$
and $m$, and $\nu=10^{-10}$, $\Omega_{e}\approx3\times10^{-4}$
: this number corresponds to low mutation regime ; $\left\langle k(0)\right\rangle /L$
is very small and individuals inside the same patch are at small genetic
distance of each other. The number of patches over which this clustering
is observable is $y^{*}\approx2.5\times10^{4}$ patches or $\approx$0.8m
in physical dimension! If we suppose that samples were taken from
few $\mu$L of bacterial intestinal residues, we see that we can expect
large neutral clustering in these samples.

The above rough estimation shows that in principle, Jeraldo et al.
data cannot exclude the neutral hypothesis for these communities. 

\section{Discussion and Conclusion.}

A fundamental question in population genetics is the number of segregating
sites in the genome of individuals of a given population submitted
to neutral mutations. We have derived in this article the probability
$u_{k}$ that two individuals, drawn at random in a population of
size $N$, differ at $k$ sites of a given gene of length $L$. This
information is usually derived in the framework of coalescent theory,
which is more general that the approach developed here. For example,
using coalescent theory, one derives the probability that $n$ individuals
drawn at random have $k$ segregating sites, while the approach here
is restricted to $n=2$, \emph{i.e.,} the \emph{pair} correlation
function. On the other hand, application of coalescent theory to \emph{finite}
sequence lengths is rather difficult and it is even harder to take
into account structured populations like geographically dispersed
individuals (see \cite{Hein2005} p45 for a detailed discussion of
these difficulties or \cite{Wakeley2013a} for recent developments
of coalescent theory). The approach we have developed here tackles
these problems rather easily, making it interesting for analyzing
more realistic systems. 

The main assumptions of the approach we have developed is that the
population is subject to \emph{neutral} selection and mutation. In
particular, we have supposed that all sites have the \emph{same} mutation
probability. This constraint can be relaxed to some extent and allow
for rate heterogeneity, a subject of intense development in the field
of molecular evolution\cite{Yang2006,Ho2014}. Consider for example
a sequence of length $L=L_{1}+L_{2}$, where sites belonging to the
$i-$th subsequence have mutation number $\Theta_{i}=L_{i}\Omega_{i}$.
Then the PGF for the whole sequence is the product of the PGF for
each sub-sequence 
\[
\phi(z)=\phi_{1}(z)\phi_{2}(z)
\]
and therefore all the moments of the whole sequence can be deduced
from the moments of the subsequences. In particular, the mean is simply
given by 
\begin{eqnarray*}
\left\langle k\right\rangle  & = & \left\langle k_{1}\right\rangle +\left\langle k_{2}\right\rangle \\
 & = & L_{1}\frac{\Omega_{1}}{1+\Omega_{1}}+L_{2}\frac{\Omega_{2}}{1+\Omega_{2}}
\end{eqnarray*}
when $\Omega_{i}\ll1$, the system behaves as having an effective
mutation rate per base equal to the weighted mean of the subsequences. 

Experimental measurement of $u_{k}$ constitutes a stringent test
of neutral theories in ecology (UNTB) introduced by Hubbell\cite{Hubbel2001}.
It has been shown by many authors\cite{Rosindell2011} that UNTB predicts
abundance distributions that are indeed observed in nature. On the
other hand, other scientists have argued that abundance distribution
is not a very selective criterion and the observed abundance patterns
in natural communities can be predicted just as well by other competing
theories \cite{Linquist2015}. The advent of modern gene sequencing
tools allows one to develop more selective criteria. O'Dwyer et al
\cite{ODwyer2015} for example have observed that the reconstructed
phylogeny of microbes from various habitats does not correspond to
that predicted by UNTB. Jeraldo et al\cite{Jeraldo2012} have used
a variant of $u_{k}$ measurement to argue against UNTB for various
microbial habitats. Their main argument that, as $u_{k}$ is a sharply
decreasing function of $k$ and is therefore in contradiction with
UNTB, is however weak. The weakness of this argument was pointed out
by D'Andrea and Ostling\cite{DAndrea2016} who showed, by combining
theoretical modeling and numerical simulation, that a neutral theory
gives rise to a $u_{k}$ function that is sharply peaked at the origin.
In this article, we have derived the explicit expression for $u_{k}$
and we show that the shape of the $u_{k}$ function depends critically
on the mutation number : it can be either peaked around the origin
or far from it, depending on the value of the mutation \emph{and}
the migration number. 

For the microbial communities discussed by Jeraldo et al\cite{Jeraldo2012},
our rough estimation suggests that the UNTB hypothesis cannot be ruled
out. Of course, the current state of DNA sequencing does not allow
for the measurement of the \emph{spatial} correlation function $u(y;k)$
for these communities. But even if we assume that the samples used
by Jeraldo et al have been mixed over few millimeters, the amplitude
of neutral clustering remains important. 

In this article, we have used the neutral WF model of competition
where all individuals compete against all the others with equal strength.
The genetic distance between individuals does not appear explicitly
in this competition. Other models can be formulated where the genetic
distance can modify the competition. In kin selection theory for example,
competition is reduced for individuals closely related to each other\cite{Gardner2011}.
On the other hand, Biancalani et al\cite{Biancalani2015}, have investigated
the problem of competitive exclusion in ecology; in their approach,
individuals are represented by a sequence which determines their use
of available ecological niches. In their model, competition is enhanced
for individuals close to each other in terms of their resource consumption,
hence giving rise to patterns of abundance in the sequence space. 

The approach we have developed here shares many limitations with other
analytical approaches: (i) As we do not track the sequences but only
pair differences, it is not possible to compute the abundance curves
as in UNTB (which, it should be stressed, was derived, only for infinite
allele models). (ii) Deriving more refined results beyond the pair
correlation function also seems problematic for the same reasons.
The simplicity of the dynamics of pair correlation function (equations
\ref{eq:fin-u0},\ref{eq:fin-uL}) is lost when $n-$correlation functions
are considered. (iii) We have considered only simple mutations. In
the field of molecular evolution however, the substitution rate of
a base in a sequence depends on the state of the base; for example,
$A\rightarrow T$ and $A\rightarrow C$ have in general different
probabilities. These different rates are contained in a substitution
matrix for which many models coexist \cite{Yang2006}. It could be
possible in principle to generalize the approach developped here to
include general substitution matrices, but the algebraic cost seems
at present to be prohibitive.

Despite all the limitations enumerated above, we believe that the
formalism developped in this article is a step forward in the search
for a better understanding of natural populations. In particular,
we are convinced that pair correlation function measurement as proposed
by Jeraldo et al will be extended to many more natural communities
and the explicit expression derived here can be used to quantify the
amplitude of neutral versus non-neutral mutations.
\begin{acknowledgments}
We thanks M. Vallade, E. Geissler, O. Rivoire and I. Junier for the
critical reading of the manuscript and fruitful discussions.
\end{acknowledgments}

\appendix

\section{The infinite allele model.\label{sec:The-infinite-allele}}

The other terms of the linear system (\ref{eq:infb-u0}-\ref{eq:infb-uk})
are obtained by the same argument as the $u_{0}(t+1)$ terms. The
probability $u_{1}(t+1)$ is obtained by considering that:
\begin{enumerate}
\item The two individuals are from the same parent (probability $a$) AND
a mutation has taken place in only one of them (probability $2\lambda(1-\lambda)$
).
\item The two individuals are from different $k=0$ parents (probability
$1-a$) AND one mutation has taken place in one of them (probability
$2\lambda(1-\lambda)$ ).
\item The two individuals are from different $k=1$ parents AND no mutation
has taken place in any of them.
\end{enumerate}
or, in other words
\begin{equation}
u_{1}(t+1)=2\lambda(1-\lambda)\left[a+(1-a)u_{0}(t)\right]+(1-\lambda)^{2}(1-a)u_{1}(t)\label{eq:inf-u1}
\end{equation}
As we will neglect terms of order $\lambda^{2}$, two individuals
at distance $k\ge2$ will not be from the same parent. Therefore the
probability $u_{k}(t)$ ($k\ge2$) is
\begin{enumerate}
\item The two individuals are from different $(k-1)$-distant parents AND
one mutation has taken place in one of them
\item The two individuals are from different $k$-distant parents AND no
mutation has taken place in either of them
\end{enumerate}
or in other words
\begin{equation}
u_{k}(t+1)=(1-a)\left[2\lambda(1-\lambda)u_{k-1}(t)+(1-\lambda)^{2}u_{k}(t)\right]\label{eq:inf-uk}
\end{equation}
These relations are summarized in the linear system (\ref{eq:infb-u0}-\ref{eq:infb-uk}).

\section{The PGF equation for $d=0$.\label{sec:The-PGF-equation.}}

We derive here the stationary PGF equation (\ref{eq:phi}) of the
main text. All notations are identical to the main text (eqs. \ref{eq:vectorial},\ref{eq:Q}).
Consider the linear form $\left<\eta\right|=(1,z,z^{2},...,z^{L})$.
By its very definition, 
\[
\left\langle \eta|u\right\rangle =\sum_{k=0}^{L}z^{k}u_{k}=\phi(z)
\]
and 
\begin{equation}
\left\langle \eta|b\right\rangle =b_{0}+b_{1}z\label{eq:etab}
\end{equation}
Moreover, straightforward matrix multiplication shows that 
\begin{equation}
\left\langle \eta|Q|u\right\rangle =z\sum_{k=0}^{L-1}A_{k}u_{k}z^{k}+\sum_{k=0}^{L}B_{k}u_{k}z^{k}+\frac{1}{z}\sum_{k=1}^{L}C_{k}u_{k}z^{k}\label{eq:etaQu}
\end{equation}
Note that $A_{L}=0$ and $C_{0}=0$; all the sums in relation (\ref{eq:etaQu})
can therefore be taken between the boundaries $0$ and $L$. Recalling
the definition of the coefficients $A_{k}$, $B_{k}$ and $C_{k}$
(eqs. \ref{eq:Ak}-\ref{eq:Bk}), we see that relation (\ref{eq:etaQu})
contains only sums of the form $\sum u_{k}z^{k}$ and $\sum ku_{k}z^{k}$.
On the other hand, 
\[
\sum_{k=0}^{L}ku_{k}z^{k}=z\phi'(z)
\]
and therefore, 
\begin{equation}
\left\langle \eta|Q|u\right\rangle =\frac{2\lambda}{L}(1-z^{2})\phi'(z)+\left(1+2\lambda(z-1)\right)\phi(z)\label{eq:eta:Q:u}
\end{equation}
The stationary solution for the probabilities $\left|u\right>$ derived
from equation (\ref{eq:vectorial}) is given by 
\begin{equation}
\left(I-(1-a)Q\right)\left|u\right>=\left|b\right>\label{eq:statu}
\end{equation}
Applying the linear form $\left<\eta\right|$ to the above relation,
we obtain
\begin{equation}
\frac{2\lambda}{L}(1-z^{2})\phi'(z)+\left(1-\frac{1}{1-a}+2\lambda(z-1)\right)\phi(z)=-b_{0}-b_{1}z\label{eq:phib}
\end{equation}
The term 
\[
1-\frac{1}{1-a}=-\frac{a}{1-a}\approx-a
\]
as we suppose that $a\ll1$ and neglect terms of $O(a^{2})$. Multiplying
both side of the relation (\ref{eq:phib}) by $L/2\lambda$ leads
to equation (\ref{eq:phi}) of the main text.

\section{The PGF equation and first moment for $d=1$.\label{sec:The-PGF-equation-space}}

The evolution equation for the spatial case (\ref{eq:uyk:0}-\ref{eq:uyk:y})
is obtained by generalizing the $d=0$ case. Consider for example
two individuals in the same patch ($y=0$). Either both of them descend
from parents of the same patch ( probability $(1-m)^{2}\approx1-2m$
) or one of them descends from a parent in a neighboring patch ( probability
$2\times m(1-m)\approx2m$ ). Distinguishing these two cases leads
to equation (\ref{eq:uyk:0}). Equations \ref{eq:uyk:1},\ref{eq:uyk:y})
are derived by following the same arguments. These equations can be
written as 
\begin{eqnarray}
\left|u(y)\right> & = & (1+m\Delta)Q\left|u(y)\right>\label{eq:uy:vectorial}\\
 & + & a\left\{ (1-2m)\delta_{y,0}+ma\delta_{y,1}\right\} \left(-Q\left|u(0)\right>+\left|b\right>\right)\nonumber 
\end{eqnarray}
The PGF is
\[
\phi(y;z)=\sum_{k=0}^{L}u(y,k)z^{k}=\left\langle \eta|u(y)\right\rangle 
\]
where $\left<\eta\right|$ was defined in the previous appendix and
is has been shown that 
\begin{eqnarray}
\left\langle \eta|Q|u(y)\right\rangle  & = & \frac{2\lambda}{L}(1-z^{2})\partial_{z}\phi(y;z)+\left(1+2\lambda(z-1)\right)\phi(y;z)\nonumber \\
 & = & {\cal L}[\phi(y,z)]\label{eq:etaQuy}
\end{eqnarray}
where the operator ${\cal L}[\phi]$ captures the right hand side
of equation (\ref{eq:etaQuy}). As $\left<\eta\right|$ and $\Delta$
commute, applying $\left<\eta\right|$ to equation (\ref{eq:uy:vectorial})
leads to 
\begin{eqnarray}
\phi(y;z) & = & (1+m\Delta){\cal L}[\phi(y,z)]\label{eq:PGFy}\\
 & + & a\left\{ (1-2m)\delta_{y,0}+ma\delta_{y,1}\right\} \left(-{\cal L}[\phi(0,z)]+f(z)\right)\nonumber 
\end{eqnarray}
where 
\[
f(z)=\left\langle \eta|b\right\rangle =a\left(1+2\lambda(z-1)\right)
\]
Note that $f(1)=a$ and $f'(1)=2\lambda a$. As ${\cal L}[\phi(y,1)]=\phi(y,1)=1$,
it can be checked that equation (\ref{eq:PGFy}) trivially verifies
$\phi(y,1)=1$. 

In order to compute $\left\langle k(y)\right\rangle $, the mean genetic
distance between two patches
\[
\left\langle k(y)\right\rangle =\sum_{k=0}^{L}ku(y,k)=\left.\frac{\partial\phi(y;z)}{\partial z}\right|_{z=1}
\]
we need to apply the projection operator $D=\left.\partial_{z}\right|_{z=1}$
to equation (\ref{eq:PGFy}). Setting $r=4\lambda/L=4\nu$ and noting
that 
\[
D{\cal L}[\phi(y,z)]=(1-r)\left\langle k(y)\right\rangle +2\lambda
\]
we find that $\left\langle k(y)\right\rangle $ must obey the relation
\begin{eqnarray*}
\left\langle k(y)\right\rangle  & = & (1+m\Delta)\left\{ (1-r)\left\langle k(y)\right\rangle +2\lambda\right\} \\
 & + & a\left\{ (1-2m)\delta_{y,0}+m\delta_{y,1}\right\} \left\{ (1-r)\left\langle k(0)\right\rangle -2\lambda(1-a)\right\} 
\end{eqnarray*}
Grouping the terms in $\left\langle k(y)\right\rangle $, dividing
by $(1-r)$ and approximating $r/(1-r)\approx r$, $(1+a)\approx1$
finally leads to equation (\ref{eq:mean:ky}) of the main text. 

The equation (\ref{eq:mean:ky}) can be solved by noting that the
solution for the bulk equation ($y\ge1$) is 
\begin{equation}
\left\langle k(y)\right\rangle =\frac{L}{2}+C\ell^{y}\label{eq:solution:ky}
\end{equation}
where $\ell$ must obey the equation 
\begin{equation}
\ell^{2}-(2+\frac{r}{m})\ell+1=0\label{eq:ell:app}
\end{equation}
this equation has two positive solutions $\ell_{1}$ and $\ell_{2}$
for which $\ell_{1}\ell_{2}=1$. As the mean distance is bounded by
$L$, the solution $\ell>1$ is unphysical and we consider only the
root $\ell<1$. Note also that $r/m=2\Omega/M$, where $\Omega$ and
$M$ are per-base mutation and migration numbers (relations \ref{eq:Omega:def},\ref{eq:migration:number}).

Equation (\ref{eq:mean:ky}) at $y=0$ provides a linear relation
between $\left\langle k(0)\right\rangle $ and $\left\langle k(1)\right\rangle $.
Using solution (\ref{eq:solution:ky}) at $y=1$ provides a second
linear relation between these two quantities. Solving this $2\times2$
linear system, we find 
\begin{equation}
\frac{\left\langle k(0\right\rangle }{L}=\frac{\Omega+M(1-\ell)}{1-2m+2\Omega+2M\left(1-\ell(1-a)\right)}\label{eq:mean:k0:solved:exact}
\end{equation}
which reduces to the expression (\ref{eq:mean:k0:solved}) for $m\ll1$
and $a\ll1$. The coefficient $C$, found from the same linear system
is 
\begin{equation}
C=(1-a)\left\langle k(0)\right\rangle -\frac{L}{2}\label{eq:C}
\end{equation}
which completes the solution. 

Equations for higher moments can be obtained by the same method and
as in the case of $d=0$, the moment of order $n$ depends only on
moments of order $n-1$. Their solution involves only straightforward,
although cumbersome algebra. As an example, consider the second factorial
moment
\[
\left\langle k_{(2)}(y)\right\rangle =\sum_{k=0}^{L}k(k-1)u(y;k)=\frac{\partial^{2}\phi(y,z)}{\partial z^{2}}
\]
which is obtained by applying the operator $D^{2}=\left.\partial_{z}^{2}\right|_{z=1}$
to equation (\ref{eq:PGFy}):
\begin{eqnarray*}
(-2r+m\Delta)\left\langle k_{2}(y)\right\rangle  & = & -Lr(1+m\Delta)\left\langle k(y)\right\rangle \\
 & + & a\left\{ (1-2m)\delta_{y,0}+m\delta_{y,1}\right\} \times\\
 &  & \left\{ \left\langle k_{(2)}(0)\right\rangle +Lr\left\langle k(0)\right\rangle \right\} 
\end{eqnarray*}
As $\left\langle k(y)\right\rangle $ is known (relation \ref{eq:solution:ky}),
this is again a solvable two term recurrence relation. 

The extension to higher dimensions is easily obtained by redefining
the discrete Laplace operator $\Delta$. For example, for $d=2$,
\begin{eqnarray*}
\Delta_{2}f(x,y) & = & -2f(x,y)\\
 &  & \frac{1}{2}\left\{ f(x+1,y)+f(x-1,y)+f(x,y-1)+f(x,y+1)\right\} 
\end{eqnarray*}
and the moments $\left\langle k(x,y)\right\rangle $ are given by
\[
\left\langle k(x,y)\right\rangle =\frac{L}{2}+C\ell^{x+y}
\]

\section{Numerical simulations.\label{sec:Numerical-simulations}}

All numerical simulations are written in C++, and data analysis is
performed by the high level language Julia\cite{Bezanson2014}. For
numerical simulation of well mixed populations (0 dimension), individuals
are represented by the binary sequence of their genome. Using sequence
lengths of powers of 2 (such as $16$ and 32) allows us to represent
each sequence as an integer, and to use the tools of the C language
to manipulate the bits directly. If for example the powers of 2 integers
are stored in an array p2, flipping the $n-$th bit of the integer
$b$ is performed by the operation b\textasciicircum{}p2{[}$n${]}
where ``\textasciicircum{}'' represents the binary XOR. 

For WF simulation of a population of size $N$, two one-dimensional
integer arrays of size $N$ are considered (present and future generations).
Each element in the second array chooses randomly a parent in the
first array and inherits its integer (genome), accompanied possibly
by a mutation with probability $\lambda$. The process is iterated
(by exchanging the role of the two arrays) $T$ times (usually $T=10N)$.
Hamming distances between all elements of the final array are computed
and stored in a new array $H$. The process is then repeated $M$
times ($M\sim10^{6}-10^{8}$) to obtain a statistically significant
array $H$, which represent the probabilities $u_{k}$. 

For one dimensional spatial simulations, the same process is applied
to two (present and future generation) two-dimensional integer arrays
B of size $Q\times N$, where $Q$ is the spatial extension of the
system and $B[q][n]$ represent the genome of individual $n$ at position
$q$. This time, each progeny chooses its parent from a neighboring
site with probability $m/2$, and from the same site with probability
$(1-m)$. 

\bibliographystyle{unsrt}

\end{document}